\documentclass[conference]{IEEEtran}
\IEEEoverridecommandlockouts
\usepackage{cite}
\usepackage{amsmath,amssymb,amsfonts}
\usepackage{algorithmic}
\usepackage{graphicx}
\usepackage{textcomp}
\usepackage{import}
\usepackage{placeins}
\usepackage{url}
\usepackage[hypertexnames=false,bookmarks=false]{hyperref} 
\hypersetup{colorlinks,linkcolor={red},citecolor={blue},urlcolor={black}} % for removing the horrible green boxes around all links 
\usepackage[euler]{textgreek} % for greek letters in text
\usepackage{blindtext}
\usepackage{ragged2e}
\usepackage[table,xcdraw]{xcolor}
\usepackage{tabularx}
\usepackage{multirow}

\newcolumntype{s}{>{\hsize=.5\hsize}X}
\newcolumntype{q}{>{\hsize=.25\hsize}X}
\newcommand{\textoverline}[1]{$\overline{\mbox{#1}}$}

\ifCLASSOPTIONcompsoc
    \usepackage[caption=false, font=normalsize, labelfont=sf, textfont=sf]{subfig}
\else
\usepackage[caption=false, font=footnotesize]{subfig}
\fi

\def\BibTeX{{\rm B\kern-.05em{\sc i\kern-.025em b}\kern-.08em
    T\kern-.1667em\lower.7ex\hbox{E}\kern-.125emX}}

\begin{document}
\bstctlcite{IEEEexample:BSTcontrol}
\addtolength{\textheight}{7.5mm}

\title{An Open-Source RRAM Compiler 
% With Author comment second vspace
\vspace{-3mm}

%Without
%\vspace{15mm}
}

%\iffalse
\author{
\IEEEauthorblockN{Dimitris Antoniadis$^*$, Andrea Mifsud$^{*\dagger}$, Peilong Feng$^{*\dagger}$, Timothy G. Constandinou$^{*\dagger\ddag}$}
\IEEEauthorblockA{$^*$Department of Electrical and Electronic Engineering, Imperial College London, SW7 2BT, UK\\$^\dagger$Centre for Bio-Inspired Technology, Institute of Biomedical Engineering, Imperial College London, SW7 2AZ, UK\\$^\ddag$Care Research \& Technology Centre, UK Dementia Research Institute, UK\\
\{dimitris.antoniadis20, a.mifsud, peilong.feng14,  t.constandinou\}@imperial.ac.uk\\
}
\vspace{-10mm}

}
%\fi 

\maketitle

\begin{abstract}
Memory compilers are necessary tools to boost the design procedure of digital circuits. However, only a few are available to academia. %Not only this, but conventional memories suffer of volatility, high power consumption, high write voltage and low endurance.
Resistive Random Access Memory (RRAM) is characterised by high density, high speed, non volatility and is a potential candidate of future digital memories. To the best of the authors' knowledge, this paper presents the first open source RRAM compiler for automatic memory generation including its peripheral circuits, verification and timing characterisation. The RRAM compiler is written with Cadence SKILL programming language and is integrated in Cadence environment. The layout verification procedure takes place in Siemens Mentor Calibre tool. The technology used by the compiler is TSMC 180nm. This paper analyses the novel results of a plethora of M x N RRAMs generated by the compiler, up to M = 128, N = 64 and word size B = 16 bits, for clock frequency equal to 12.5\,MHz. Finally, the compiler achieves density of up to 0.024\,Mb/mm\textsuperscript{2}.

\end{abstract}
\section{Introduction} \label{sec:Introduction}

% REMOVE CITATION TO MY SELF
{
%The most essential and vital components of a digital system are the Random Access Memory (RAM) and the Central Processing Unit (CPU)~\cite{antoniadis2021open}. 
}
The progress of the technology has lead to an aggressive scaling trend of Integrated Circuits (IC), generating great demand for high density, high speed and low power memories~\cite{chen2020reram}. However, customised memory design is a resource-intensive task~\cite{xu2007flexible,guthaus2016openram,shah2010fabmem}. Therefore, a number of memory compilers has been presented before~\cite{shah2010fabmem,xu2007flexible,guthaus2016openram,wu201065nm, goldman2014synopsys, clinton20185ghz}. These compilers mainly focus on two types of memories, the volatile memories (e.g. static or dynamic random access memory) and the non volatile memories (e.g. flash memory)~\cite{chen2016review,zahoor2020resistive}. Volatile memories suffer of high power consumption and loss of data on power off, while non volatile memories require high write voltage and suffer of low endurance and low speed~\cite{zahoor2020resistive,maheshwari_hybrid_2020}.

A novel emerging non volatile memory which is characterised by scalability, high speed, high density and low power is the Resistive Random Access Memory (RRAM). RRAM uses memristors in its memory cells~\cite{strukov2008missing,chen2016review,maheshwari_hybrid_2020,yang2013memristive,zahoor2020resistive, ielmini2018memory, stathopoulos2019electrical}. Therefore, a RRAM compiler is an essential tool to boost the RRAM design procedure and facilitate the research of the memristive device properties~\cite{antoniadis2021open}.

The work presented in \cite{antoniadis2021open} described a novel RRAM architecture and demonstrated the results only of the RRAM array generation and its layout verification. This paper extends this work by improving the RRAM architecture described in \cite{antoniadis2021open} and presents the first open source RRAM compiler which automatically generates the RRAM including its peripheral circuits, verifies their layout and performs timing characterisation. The features of the RRAM compiler presented in this work are presented in Tab.~\ref{tab:version_comparison}. It is clear that the work presented in ~\cite{antoniadis2021open} includes partially only two features of the RRAM compiler presented in this paper. The proposed RRAM compiler includes new features regarding Place and Route (P\&R) of memory blocks, co-integration of analogue and digital blocks, layout verification of the whole RRAM and timing characterisation.  The source code of the proposed RRAM compiler has been published on Github:

\begin{center}
    \url{https://github.com/akdimitri/RRAM_COMPILER}
\end{center}

This Section briefly described the background regarding RRAMs, memory compilers and the new features of the proposed RRAM compiler. Section \ref{sec:RRAM_architecture} presents the RRAM architecture and its most important blocks. In Section \ref{sec:RRAM_Compiler} the RRAM compiler structure and the characterisation procedure are described. Section \ref{sec:results} analyses the results produced by a number of automatically generated designs. Finally, Section \ref{sec:conclusion} summarises this paper.

\begin{figure}[!t]
\centerline{\includegraphics[width=\columnwidth]{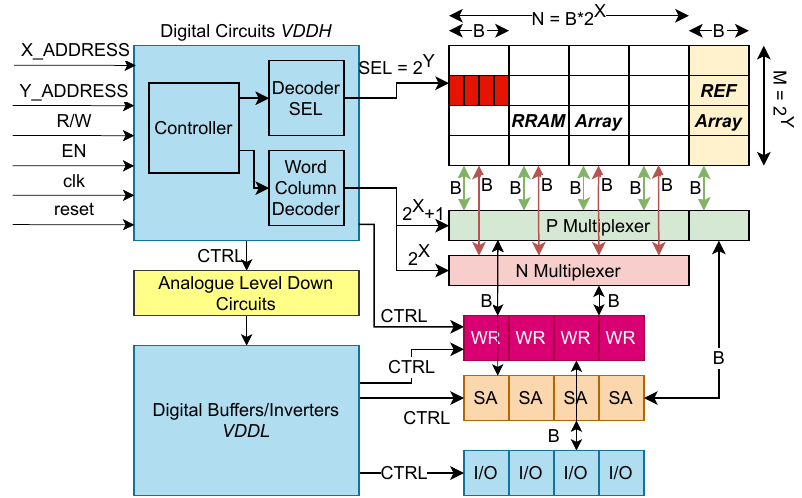}}
%\vspace{-3mm}
\caption{Simplified block diagram of the RRAM architecture. The size of the RRAM array is M\,$\times$\,N, where M = 2\textsuperscript{Y} and N = B2\textsuperscript{X}. Parameter M is the number of rows and parameter N is the number of columns. Parameter B is the length of the memory words. Figure adapted from Fig. 1. in Ref. \cite{antoniadis2021open}.}
\label{fig:RRAM_architecture}
\vspace{-3mm}
\end{figure}

\begin{table}[t!]
\caption{Open-source RRAM Compilers comparison between this work and work presented in \cite{antoniadis2021open}.
}
\label{tab:version_comparison}
\vspace{-2mm}
\resizebox{\columnwidth}{!}{
\centering
\begin{tabular}{ll|cc}
\hline
Architecture                                   & Functionalities                                  & Work \cite{antoniadis2021open} & This Work \\ \hline
RRAM array                                     & RRAM array generation                            & Yes  & Yes       \\ \hline
\multirow{5}{*}{Peripheral analogue circuits} 			& Reference cells generation                       & No   & Yes       \\
                                               & Multiplexers generation                          & No   & Yes       \\
                                               & Write Amplifiers generation                & No   & Yes       \\
                                               & Analogue P\&R					                  & No   & Yes       \\
                                               & Analogue co-integration							  & No   & Yes       \\ \hline
\multirow{4}{*}{Peripheral digital circuits}   			& Synthesis                      					& No   & Yes       \\
												& Implementation                     				& No   & Yes       \\
                                               & Design constraints generation            & No   & Yes       \\
                                               & GDS and Verilog import                           & No   & Yes       \\ \hline
\multirow{5}{*}{System-level}					& Mixed-signal P\&R 					  				& No   & Yes       \\
												& Mixed-signal co-integration 						& No   & Yes       \\
                                               & Layout generation                                & No  & Yes       \\
                                               & Layout verification                              & Yes   & Yes       \\
                                               & System characterisation                          & No   & Yes       \\ \hline
\end{tabular}
}
\vspace{-3mm}
\end{table}
\section{RRAM Architecture} \label{sec:RRAM_architecture}
\subsection{RRAM Architecture Overview}

A simplified block diagram of the proposed RRAM architecture is shown in Fig.~\ref{fig:RRAM_architecture}. The size of the RRAM array is M\,$\times$\,N, where M = 2\textsuperscript{Y} and N = B2\textsuperscript{X}. A word of B memory cells is shown under red colour. Blue blocks represent digital circuits, while the rest of them represent analogue circuits. Compared to the architecture presented in \cite{antoniadis2021open}, reference cells array, B additional of P multiplexer switches and level-shifters between digital blocks have been introduced. 

 A multiplexer switch consists of a grounding NMOS and a transmission gate. The P multiplexer has one more block of B switches compared to the N multiplexer due to the fact that it connects the reference cells array to the sense amplifiers.

\begin{figure}[!t]
\centerline{\includegraphics[width=\columnwidth]{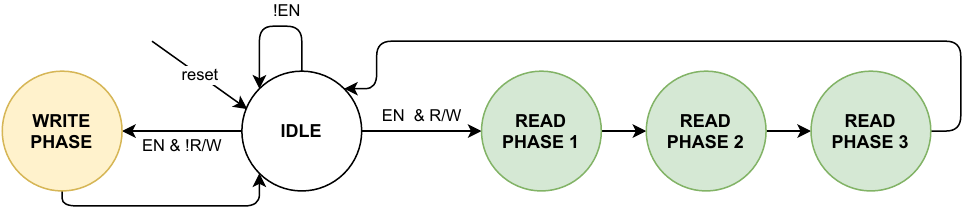}}
\vspace{-4mm}
\caption{Simplified finite state machine of the controller of the RRAM with respect to the circuits of Fig.~\ref{fig:RRAM_architecture}. If EN and \textoverline{RW} are received, then the controller generates necessary signals to execute write operation. On the other hand, if EN and RW is received, read operation begins. Read procedure is completed in three phases based on the three-phase operation of the sense amplifier.}
\label{fig:FSM}
\vspace{-1mm}
\end{figure}

Six signals are needed to control the RRAM in contrast with \cite{antoniadis2021open}, where only the three very basic were introduced. These are the X\_ADDRESS[X:1] signal (address of the desired word column), the  Y\_ADDRESS[Y:1] signal (address of the desired row), the EN signal (enable chip), the R/W signal (read or write operation), the clock of the chip and the reset signal. The desired word to be read or written is interfaced through I/O ports. A simplified Finite State Machine (FSM) of the digital controller is depicted in Fig.~\ref{fig:FSM}. 

RRAM uses three power supplies. Memory cells have been designed with high voltage (\emph{VDDH}) technology transistors. Therefore, a number of digital circuits, P and N multiplexer also operate under \emph{VDDH} voltage. On the other hand sense amplifiers, have to ensure that the voltage applied on the bitlines does not alter the state of the memristor of the memory cells, thus \emph{VDDL} has been chosen for the sense amplifiers~\cite{antoniadis2021open}. The sense amplifier incorporates low voltage (\emph{VDDL}) technology transistors. The digital signals that control (CTRL) \emph{VDDL} circuits have to be levelled down through an analogue level-shifter circuit and they are propagated to the sense amplifiers array and I/O ports through digital buffers or inverters. In order to provide flexibility on the write voltage, a third power supply \emph{VDDW} has also been included in the design.

The reference cell consists of a NMOS transistor. The drain source resistance of the transistor represents the reference resistance R\textsubscript{REF}. Its resistance is compared against the corresponding memory cell resistance, when it is sensed by the corresponding sense amplifier. The reference cells are integrated alongside the RRAM array, as they share same lines and have same layout size. The size of the reference cells array is M\,$\times$\,B.

\subsection{Write Circuit}
\begin{figure}[!t]
\centerline{\includegraphics[width=\columnwidth]{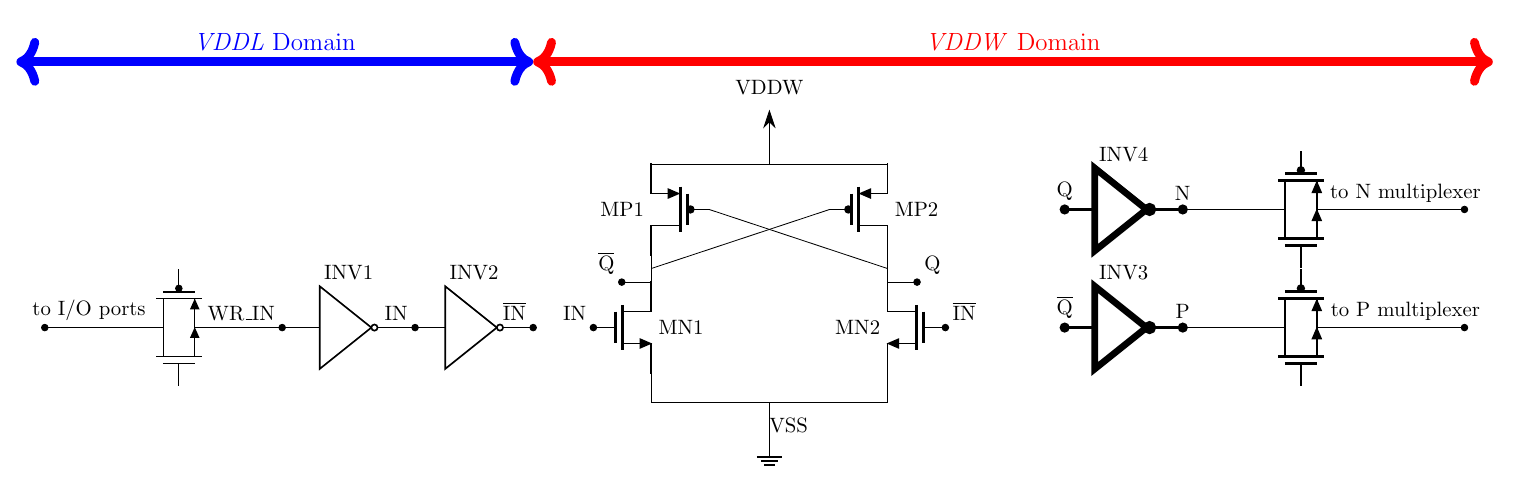}}
\vspace{-5mm}
\caption{Write circuit. The input signal at I/O port is a value of 0 or \emph{VDDL}, thus this value has to be levelled up to \emph{VDDW} and then to be propagated correctly to the corresponding P and N lines.}
\label{fig:write_circuit}
\vspace{-6mm}
\end{figure}

% REMOVE IT AS REDUNDANT
%All circuits presented so far, operate under \emph{VDDH} voltage. It was mentioned above that a high voltage should be applied on P or N in order to write the desired state at the memory. 
The write voltage (\emph{VDDW}) differs based on the memristor materials. %Therefore, a part of the write circuit has to operates under \emph{VDDW} power supply in order to use the correct write voltage during write procedure.
The write circuit is shown in Fig.~\ref{fig:write_circuit}. The devices in Figures with thick line make use of transistor models able to support up to \emph{VDDH} voltage, while the others are able to support up to \emph{VDDL} respectively. WR\_IN can be 0 V or \emph{VDDL}. Transistors MN1, MN2, MP1, MP2 and INV2 form a conventional level-shifter with positive feedback\cite{dwivedi2012level}. Thus, if WR\_IN is digital 0, which is modelled in this implementation as Low Resistance State (LRS), then Q will go high, as such, P will be  \emph{VDDW}  and N will be 0\,V. On the contrary, if WR\_IN is digital 1 or \emph{VDDL}, which is modelled in this implementation as High Resistance State (HRS), then P will be 0\,V and N will be \emph{VDDW}. These values will be propagated to the desired memory cell through the P and N multiplexers.

\subsection{Sense Amplifier}
\begin{figure}[!t]
\centerline{\includegraphics[width=\columnwidth]{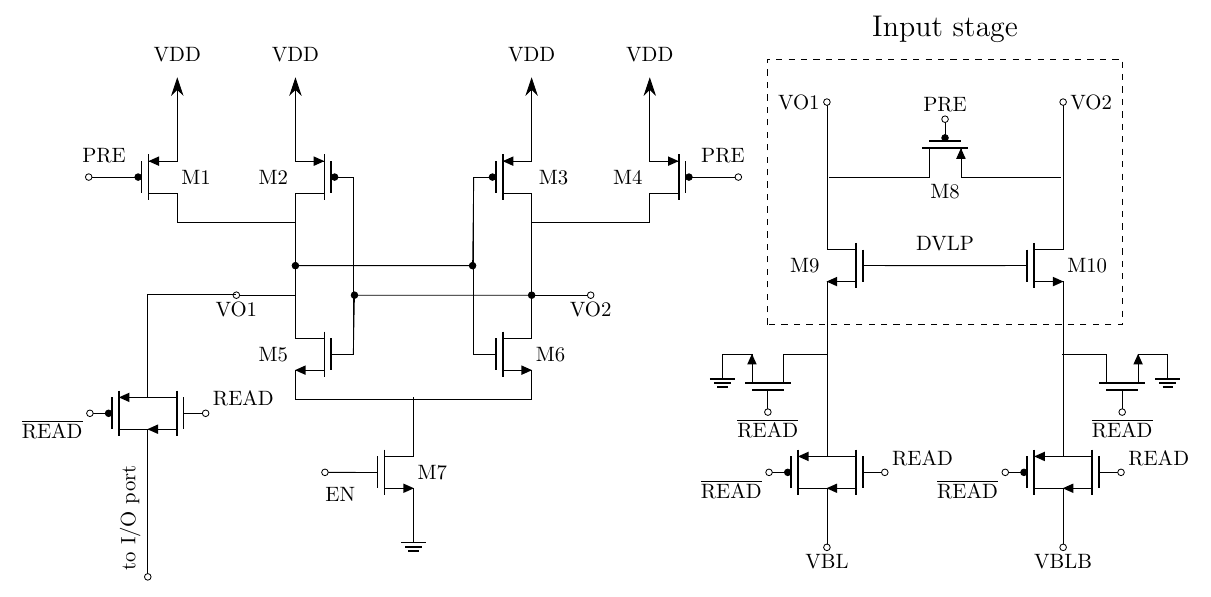}}
\vspace{-3mm}
\caption{Sense amplifier circuit. On the left side, a typical cross coupled sense amplifier is shown. The input stage is located on the right. Low impedance of NMOS transistor is used to achieve fast read operation and avoid large loading time duration on bitlines.}
\label{fig:sense_circuit}
\vspace{-6mm}
\end{figure}

The sense amplifier is based on those presented in~\cite{tolic2019design, suma2017analysis} and it is shown in Fig.~\ref{fig:sense_circuit}. Transistors M2, M5, M3, M6 form a typical cross coupled inverters sense amplifier. Transistors M1 and M4 precharge VO1, VO2 to \emph{VDD}. In this case \emph{VDD} refers to \emph{VDDL}. A common gate configuration is used for the input stage, keeping the input impedance (M9 and M10 source terminal) as low as possible. Transistors M9 and M10 operate in triode region and they force a low voltage on bitlines VBL, VBLB. Transistor M8 is an equaliser that ensures that equal values are set on VO1 and VO2. Three additional transistors forming a transmission gate and a grounding NMOS, enable the sense of the memory cell when read operation is executed. Further to this, there is also a transmission gate that connects VO1 to I/O ports.

The sense amplifier operates in three phases. Its operation is illustrated in Fig.~\ref{fig:rram_read_test}. When READ signal is high, phase 1 starts. Signal PRE remains low, DVLP goes high and  current is drawn from the memory cell and the reference cell. In phase 2, PRE goes high, therefore, VO1 and VO2 are no longer equal and the voltage difference, which is developed between them, is rapidly enhanced. In phase 3, DVLP goes low and EN (or EN\_SA) goes high, forcing the result to rail to rail output voltage value. In Fig.~\ref{fig:rram_read_test}, vertical lines READ TEST 1 and READ TEST 2  show the moment where EN\_SA is set to high and the output of the first (Z\_SA[0]) and the second (Z\_SA[1]) sense amplifier has been propagated to the bus (Z\_BUS[3:0]) of the tri state buffers (I/O ports).

%\subsection{Digital Controller of RRAM}

%In order for the analogue circuits to be controlled, a digital controller is needed. A simplified Finite State Machine (FSM) of the controller is depicted in Fig.~\ref{fig:FSM}. The three read phases are shown under green colour, while write phase is shown under yellow colour. During both read and write operations, CTRL signals (READ, WRITE, DVLP, PRE, EN\_SA) are set accordingly. Additionally, the decoders are enabled to choose the correct word from memory.

% REMOVED AS REDUNDANT
{
%A number of those control signals have to be levelled down to \emph{VDDL} in order to control \emph{VDDL} analogue circuits. Therefore, analogue level down buffers were created. These buffers are shown in blue colour in Fig.~\ref{fig:RRAM_architecture}, between the two power domain digital circuits. The levelled down signals drive the desired analogue blocks. This decision was taken because the driving strength of a digital circuit can be defined in the tcl constraints script, while an analogue implementation would require parameterisation of analogue blocks leading to time consuming and less robust implementation.
}

\section{RRAM compiler} \label{sec:RRAM_Compiler}
%\subsection{RRAM compiler Overview}

% REMOVE AS REDUNDANT
%The RRAM compiler is directly integrated within the Cadence Virtuoso environment. A plethora of SKILL programming language scripts comprise the RRAM compiler~\cite{skillRef,skilluser}. 
% REMOVED AS REDUNDANT 
%A very simplified procedure of the RRAM compiler is shown on Fig.~\ref{fig:flowchart}. 

\begin{figure}[!t]
%\vspace{-6mm}
\centerline{\includegraphics[width=\columnwidth, trim = 8 8 8 8, clip]{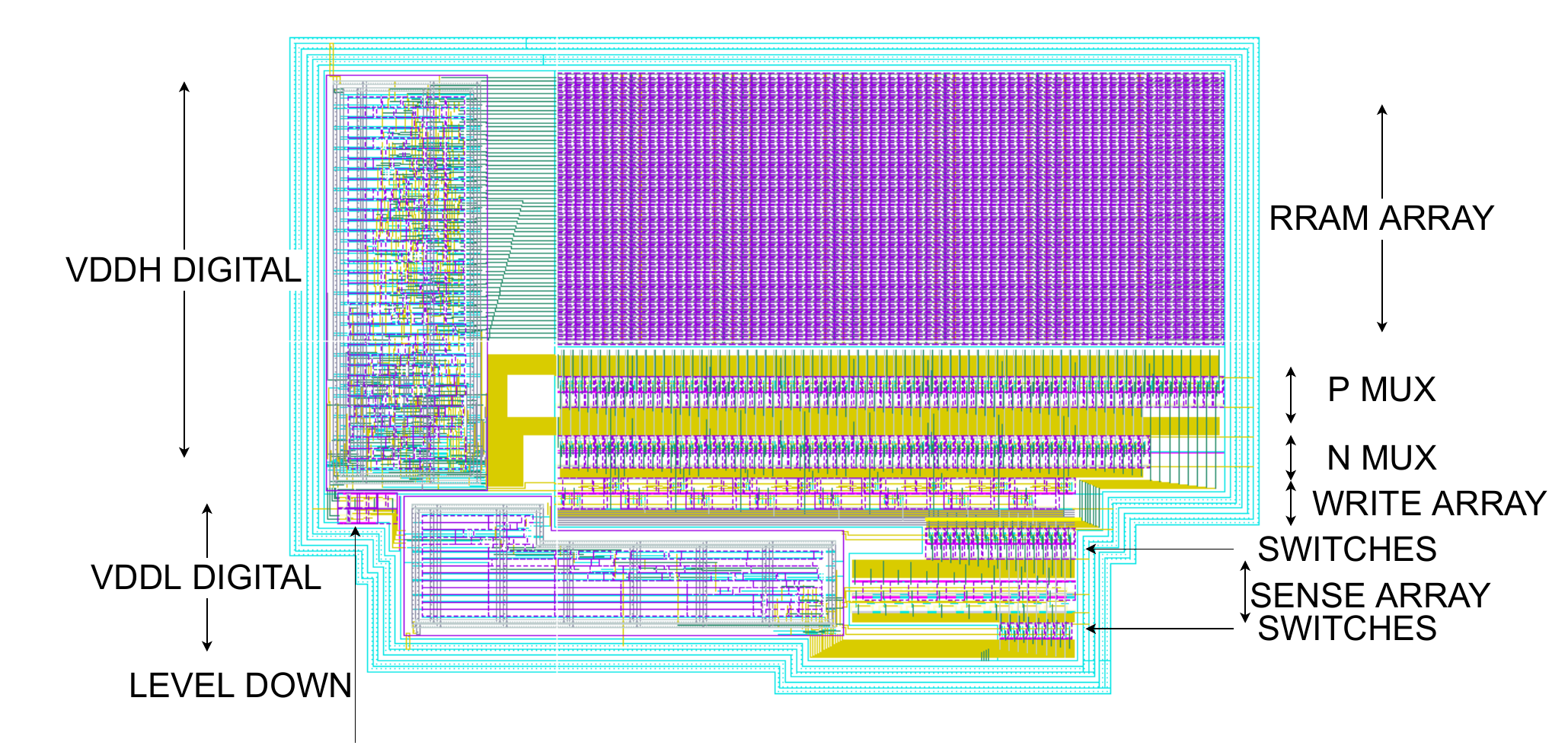}}
\vspace{-3mm}
\caption{Generated layout by RRAM compiler. The dimensions of this circuit are M = 64, N = 64 and B = 8. Its size is 524.3\,um\,$\times$\,353.5\,um.}
\label{fig:rram_layout}
\vspace{-6mm}
\end{figure}

The RRAM compiler is written in SKILL language and is invoked by Command Interpreter Window (CIW) of Cadence Virtuoso 6.1.8~\cite{skillRef,skilluser}. The very basic arguments are the dimensions (M, N, B) of the memory. Further optional arguments, such as the minimum clock frequency can be specified. The RRAM compiler procedure follows the features presented from top to bottom in Tab.~\ref{tab:version_comparison}. The RRAM compiler makes use of Cadence Virtuoso for analogue design and Cadence Genus and Innnovus for digital design~\cite{virtuosoadexluser,virtuosolayoutuser,virtuososchuser,virtuosoxpluser,genususer,innovususer}. It also verifies the layout correctness of the generated RRAMs by executing Design Rule Checking (DRC) and Layout Versus Schematic (LVS) tests in Mentor Calibre tool~\cite{calibreinter,calibreuser}. The post-layout view of the RRAM generated by Parasitic Extraction (PEX) tool of Mentor Calibre is used in the timing characterisation procedure. The timing characterisation procedure is invoked by an automatically generated OCEAN script~\cite{ocean}. An example of an automatically generated layout by the RRAM compiler is shown in Fig.~\ref{fig:rram_layout}.

% REMOVED AS REDUNDANT
%Then, by using the pre-designed base cells (memory cell, sense amplifier, write circuit, logic gates etc.), the schematic and the layout of the analogue part of the RRAM is generated, verified and its post layout view is extracted~\cite{antoniadis2021open,virtuosoadexluser,virtuosolayoutuser,virtuososchuser,virtuosoxpluser,calibreinter,calibreuser}. 

% REMOVED AS REDUNDANT
{
%The layout implementation undergoes Design Rule Checking (DRC) and Layout Versus Schematic (LVS) tests~\cite{calibreinter,calibreuser}. Due to the fact that all blocks have been designed with horizontal dimensions multiple of 5\,um, clear placement and routing patterns are achieved.

% REMOVED AS IT WAS DESCRIBED IN TABLE
%Based on the provided arguments and the size of the analogue part, the necessary verilog, tcl and other files are generated for the digital implementation procedure. The compiler invokes Cadence Genus and subsequently Cadence Innovus to design both \emph{VDDL} and \emph{VDDH} digital circuits~\cite{genususer,innovususer}. %Both pre-synthesis and post-synthesis simulations with Standard Delay Format (SDF) back annotation are executed, to ensure the correctness of the designs.The digital designs (GDS, verilog) are finally imported to Cadence Virtuoso.

% REMOVED AS IT WAS DESCRIBED IN TABLE
%Lastly, the compiler places automatically, routes the top level design of the desired RRAM and verifies its layout correctness by executing Design Rule Checking (DRC) and Layout Versus Schematic (LVS) tests in Mentor Calibre tool~\cite{calibreinter,calibreuser}. Additionally, a testbench is created to automatically characterise the RRAM. The characterisation procedure is invoked by an automatically generated OCEAN script~\cite{ocean}.
}

%An example of an automatically generated layout by the RRAM compiler is shown in Fig.~\ref{fig:rram_layout}. %The size of this RRAM is 64\,$\times$\,64 and B = 8. The density is 0.021 Mb/mm\textsuperscript{2}.

% REMOVED AS REDUNDANT
{
% \begin{figure}[!t]
% \centerline{\includegraphics[width=\columnwidth]{Fig/rram_compiler_main.drawio.pdf}}
% \vspace{-3mm}
% \caption{Simplified RRAM compiler flowchart. Under green colour, procedures that are invoked in Cadence Virtuoso, under red colour procedures executed by Siemens Mentor Calibre and under yellow colour, procedures executed by the Cadence digital environment (Genus/Innovus).}
% \label{fig:flowchart}
% \vspace{-6mm}
% \end{figure}
}

The RRAM characterisation procedure generates a testbench which includes the RRAM block and a verilog block with test signals. The post-layout view of the analogue part of the RRAM and the  P\&R verilog netlist of the digital circuits are used by the mixed signal simulator in Cadence Virtuoso. The automatically generated RRAMs are tested for Typical Typical (TT), Fast Slow (FS), Slow Fast (SF), Slow Slow (SS) and Fast Fast (FF) NMOS and PMOS models at nominal corner. During the simulation, the Standard Delay Format (SDF) files produced by Innovus after P\&R are used to provide realistic results for the digital circuits.  The testbench performs initially a reset and then two extreme conditions write (W1, W2) and two read (R1, R2) tests.

% REMOVED TO SAVE SPACE
{
% \begin{figure}[!t]
% \centerline{\includegraphics[width=\columnwidth,trim=4 4 4 4,clip]{Fig/write_test.png}}
% \vspace{-3mm}
% \caption{Write tests W1 and W2 for a memory with M = 32, N = 32, B = 4 and clock period equal to 80\,ns. It should be noted that the signals shown here start counting by zero (0). The vertical lines, show the moment when checks are performed. The addresses of the word under test will be in this case X\_ADDRESS = 2\textsuperscript{X}-1-1 = 2\textsuperscript{3}-1-1 = \textsubscript{(10)}6 = \textsubscript{(2)}110 and  Y\_ADDRESS = 2\textsuperscript{Y}-1 = 2\textsuperscript{5}-1 = \textsubscript{(10)}31 = \textsubscript{(2)}11111.} 
% \label{fig:rram_write_test}
% \vspace{-6mm}
% \end{figure}
}

By inspection of Fig.~\ref{fig:rram_layout}, it is clear that the worst case read/write operation occurs for the words that are located at the top right corner of the layout, where is the furthest distance from the read/write circuits. Therefore, the loading time of the lines involved in read/write operation for these cells will be the greatest. In the first write test (W1), the word is located at position X\_ADDRESS = 2\textsuperscript{X}-1 and  Y\_ADDRESS = 2\textsuperscript{Y}. This word is the second to the last word on the top row of the RRAM array.  The LRS and HRS values depend on the material characteristics. It is well known that a number of memristors get resistance values in the range of a few Ohms to tens of kOhms~\cite{stathopoulos2019electrical,nikiruy2018precise}. Given that, the memristance of the memory cells is emulated with a resistance equal to 1\,MΩ as an extreme value to examine worst case settling time across the memory cell. The desired value to be written on this test is {10...10}, resulting in V\textsubscript{PN} (Voltage across P and N terminal of memory cell) being positive for a 0, and negative for a 1. The voltage across P and N terminals of the word is checked  at 0.4\,$\times$\,clock time later than the write phase has started by the controller. This is because the clock waveform that is used by the synthesis has 50\% duty cycle. Therefore, this value has to be set up in time. The test checks whether, at this point of time, the desired V\textsubscript{PN} is greater than 0.7\,$\times$\,\emph{VDDW} when I/O = 0 or V\textsubscript{PN} is less than -0.7\,$\times$\,\emph{VDDW} when I/O = 1. As a default value, \emph{VDDW} is set by the compiler to 3.3\,V. Memristors write voltage can vary from around 2\,V to up to 10\,V or more~\cite{stathopoulos2019electrical}. The limit 0.7\,$\times$\,\emph{VDDW} is a demonstration limit and should be set according to the device characteristics. The W2 test performs the same checks. The only difference is that the desired value to be written is the complementary one {01...01} compared to the previous test. %An example of the W1, W2 tests for the first two bits of the word under test in the memory is shown in Fig.~\ref{fig:rram_write_test}.

\begin{figure}[!t]
%\vspace{-6mm}
\centerline{\includegraphics[width=\columnwidth,trim=4 4 4 4,clip]{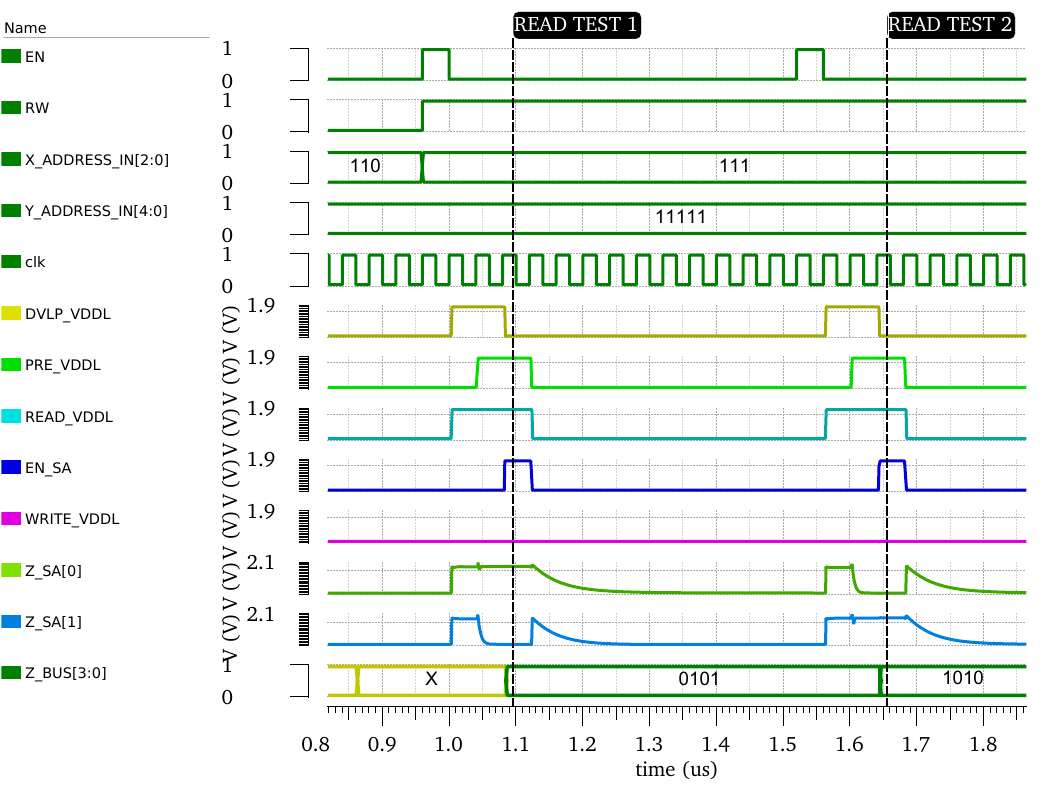}}
\vspace{-4mm}
\caption{Read tests R1 and R2 for a memory with M = 32, N = 32, B = 4 and clock period equal to 40\,ns. It should be noted that the signals shown here start counting by zero (0). The vertical lines, show the moment when checks are performed. The addresses of the word under test will be in this case X\_ADDRESS = 2\textsuperscript{X}-1 = 2\textsuperscript{3}-1= \textsubscript{(10)}7 = \textsubscript{(2)}111 and  Y\_ADDRESS = 2\textsuperscript{Y}-1 = 2\textsuperscript{5}-1 =   \textsubscript{(10)}31 = \textsubscript{(2)}11111. Signals Z\_SA are the VO1 signals with respect to Fig.~\ref{fig:sense_circuit} of the corresponding sense amplifiers. Signal Z\_BUS is output of the Tri State Buffer of the RRAM memory. The rest of the signals are the digital control signals that have been levelled down.} 
\label{fig:rram_read_test}
\vspace{-6mm}
\end{figure}

Similarly, two tests are executed to verify correctness of the read operation. In read tests, the last word of the top row is selected. The address of this word is X\_ADDRESS = 2\textsuperscript{X} and  Y\_ADDRESS = 2\textsuperscript{Y}. In this case, the memristance of each memory cell is also emulated with a resistor (representing extreme HRS and LRS values). In the first read (R1) test, the resistances of the cells of the word under test are initialised with respect to a value $a < 1$. By default $a$ is set equal to 0.3. Then, LRS is equal to $a\cdot$R\textsubscript{REF} and HRS is equal to $a^{-1}\cdot$R\textsubscript{REF}. In this case, R\textsubscript{REF} was set equal to 32.5\,kΩ as a typical median resistance value of a memristor. In the first read test (R1), the word is initialised to value {LRS,\,HRS...LRS,\,HRS}. Therefore, for HRS, logical 1 should be read at the output 1 and for LRS, logical 0 respectively. Same as previously, the check takes place 0.4\,$\times$\,clock time later than the third phase of the read procedure, as the read value has to be set up in time for the tri state buffers to hold it. In this case, regarding the cells that have been initialised with the HRS value, it is checked whether the value at the output of the sense amplifier is greater than 0.83\,$\times$\,\emph{VDDL}. For those which have been initialised with the LRS value, it is checked if the value at the output of the sense amplifier is less than 0.16\,$\times$\,\emph{VDDL}. The second read test is the same as above. The only difference is that the complementary value {HRS,\,LRS...HRS,\,LRS} is set on the word under test. An example of the R1, R2 tests for the first two bits of the word under test in the memory is shown in Fig.~\ref{fig:rram_read_test}.

\section{Results} \label{sec:results}

Multiple sizes were tested, as it is shown on Tab.~\ref{tab:results_80n}. Maximum clock frequency was set to 25\,MHz, as this is the maximum clock frequency that meets the digital synthesis design constraints.  A table with the results for 12.5\,MHz is shown on Tab.~\ref{tab:results_80n}. Memories succeed in all write tests, while some of them fail to pass the read tests mainly for FS and FF corners. This occurs due to the fact that only one output of the  sense amplifiers is used in this topology. This output is connected to a bus while the other one is floating, as such, there is an imbalance between the driving output load. The author suggests that the sense amplifiers should be isolated by connecting their input/outptus to buffers. Write operations are correct (all corners) for an up to 8\,kb memory and write time 80\,ns. Write test, in greater frequencies, may fail due to the big loading time on the terminals of the memory cells. Additionally, read operations are correct (all corners) for an up to 8\,kbit memory with B = 4 and access time approximately equal to 2/3\,clock, which is 160\,ns. Fastest access time is equal to 80\,ns and it is achieved for an up to 1\,kbit memory with B = 4 and clock frequency 25\,MHz. The best density (RRAM bit size / RRAM layout size) achieved by the compiler is 0.024\,Mb/mm\textsuperscript{2} in TSMC 180\,nm technology.

% This was unavoidable and removed as less important mistake
%The RRAM compiler achieves suboptimal optimisation as there are empty spots in the layout and a couple digital control lines are polygonal. This is because the technology used for this implementation forces the digital output lines to have 0.56\,um or multiple of 0.56\,um pitch. On the other hand analogue lines need to have greater pitch in order not to violate DRC test. Future work should focus on addressing this problem. A potential solution would be, the memory cells of the memory to be redesigned with vertical pitch multiple of 0.56\,um and the control lines on the bottom of the design could use  2\,$\times$\,0.56\,um pitch as an unavoidable trade off.

Furthermore, by using three power supplies the design becomes less robust and complex. The author suggests that the architecture could be redesigned and simplified by using only one power supply \emph{VDDH}. The write circuit could be replaced with a programmable digital to analogue converter (DAC) to provide the necessary flexibility on the write voltage. Finally, the reference array could be potentially replaced by a programmable current mirror.

 Memristance is a property that arises particularly in thin film devices, which are extremely small in size. The current implementation takes into account their size by placing corresponding pins on the top metals of RRAM cells. In this way, memristors can be placed on top of the RRAM array after the CMOS fabrication of the memory. However, the RRAM compiler uses ideal resistors for their emulation in the timing characterisation procedure. As a future improvement, the authors suggest the use of a realistic model of a memristor in order to account for even more realistic results. Additionally, temperature variation should be included in the tests. Furthermore, the power estimation of the RRAM should be introduced in future versions of the RRAM compiler.

{\renewcommand{\arraystretch}{1}
\begin{table}[!t]
%\vspace{-6mm}
\caption{Results produced by memory characterisation procedure with clock $80\,$ns (12.5\,MHz). 
%Under red colour tests that failed on READ1 (R1), READ2 (R2), WRITE1 (W1) or WRITE2 (W2).
}
\label{tab:results_80n}
\vspace{-3mm}
\resizebox{\columnwidth}{!}{
\centering
%\Small
%\footnotesize
\large
\begin{tabularx}{\textwidth}{|q|q|q|
>{\columncolor[HTML]{Abfeb2}}X |
>{\columncolor[HTML]{Feb3ab}}X |
>{\columncolor[HTML]{Abfeb2}}X |
>{\columncolor[HTML]{Abfeb2}}X |
>{\columncolor[HTML]{Feb3ab}}X |s|}
\hline
\cellcolor[HTML]{C0C0C0}N & \cellcolor[HTML]{C0C0C0}M & \cellcolor[HTML]{C0C0C0}B & \cellcolor[HTML]{C0C0C0}Nominal       & \cellcolor[HTML]{C0C0C0}FS   & \cellcolor[HTML]{C0C0C0}SF            & \cellcolor[HTML]{C0C0C0}SS                   & \cellcolor[HTML]{C0C0C0}FF   & \cellcolor[HTML]{C0C0C0}report                                                                                                                                            \\ \hline
32                        & 32                        & 4                         & pass                                  & \cellcolor[HTML]{Abfeb2}pass & pass                                  & pass                                         & \cellcolor[HTML]{Abfeb2}pass & \href{https://rawcdn.githack.com/akdimitri/RRAM_COMPILER/24f024d7e8ab4d575474f9ea0d02535fe02467d2/raw/CHARACTERISATION/32_32_4/80_TB_TOP_32_32_4.html}{\textit{link}}     \\ \hline
64                        & 16                        & 4                         & pass                                  & \cellcolor[HTML]{Abfeb2}pass & pass                                  & pass                                         & \cellcolor[HTML]{Abfeb2}pass & \href{https://rawcdn.githack.com/akdimitri/RRAM_COMPILER/24f024d7e8ab4d575474f9ea0d02535fe02467d2/raw/CHARACTERISATION/64_16_4/80_TB_TOP_64_16_4.html}{\textit{link}}     \\ \hline
64                        & 16                        & 8                         & \cellcolor[HTML]{Feb3ab}fail (R1)     & fail (R1, R2)                & pass                                  & pass                                         & fail (R1, R2)                & \href{https://rawcdn.githack.com/akdimitri/RRAM_COMPILER/24f024d7e8ab4d575474f9ea0d02535fe02467d2/raw/CHARACTERISATION/64_16_8/80_TB_TOP_64_16_8.html}{\textit{link}}     \\ \hline
64                        & 32                        & 4                         & pass                                  & \cellcolor[HTML]{Abfeb2}pass & pass                                  & pass                                         & \cellcolor[HTML]{Abfeb2}pass & \href{https://rawcdn.githack.com/akdimitri/RRAM_COMPILER/24f024d7e8ab4d575474f9ea0d02535fe02467d2/raw/CHARACTERISATION/64_32_4/80_TB_TOP_64_32_4.html}{\textit{link}}     \\ \hline
64                        & 32                        & 8                         & \cellcolor[HTML]{Feb3ab}fail (R1, R2) & fail (R1, R2)                & pass                                  & pass                                         & fail (R1, R2)                & \href{https://rawcdn.githack.com/akdimitri/RRAM_COMPILER/24f024d7e8ab4d575474f9ea0d02535fe02467d2/raw/CHARACTERISATION/64_32_8/80_TB_TOP_64_32_8.html}{\textit{link}}     \\ \hline
64                        & 64                        & 4                         & pass                                  & \cellcolor[HTML]{Abfeb2}pass & pass                                  & pass                                         & \cellcolor[HTML]{Abfeb2}pass & \href{https://rawcdn.githack.com/akdimitri/RRAM_COMPILER/24f024d7e8ab4d575474f9ea0d02535fe02467d2/raw/CHARACTERISATION/64_64_4/80_TB_TOP_64_64_4.html}{\textit{link}}     \\ \hline
64                        & 64                        & 8                         & pass                                  & fail (R1, R2)                & pass                                  & \cellcolor[HTML]{FFCC67}near                 & fail (R1, R2)                & \href{https://rawcdn.githack.com/akdimitri/RRAM_COMPILER/24f024d7e8ab4d575474f9ea0d02535fe02467d2/raw/CHARACTERISATION/64_64_8/80_TB_TOP_64_64_8.html}{\textit{link}}     \\ \hline
128                       & 16                        & 4                         & pass                                  & \cellcolor[HTML]{Abfeb2}pass & pass                                  & pass                                         & \cellcolor[HTML]{Abfeb2}pass & \href{https://rawcdn.githack.com/akdimitri/RRAM_COMPILER/24f024d7e8ab4d575474f9ea0d02535fe02467d2/raw/CHARACTERISATION/128_16_4/80_TB_TOP_128_16_4.html}{\textit{link}}   \\ \hline
128                       & 16                        & 8                         & pass                                  & fail (R1, R2)                & pass                                  & pass                                         & fail (R1, R2)                & \href{https://rawcdn.githack.com/akdimitri/RRAM_COMPILER/24f024d7e8ab4d575474f9ea0d02535fe02467d2/raw/CHARACTERISATION/128_16_8/80_TB_TOP_128_16_8.html}{\textit{link}}   \\ \hline
128                       & 16                        & 16                        & \cellcolor[HTML]{Feb3ab}fail (R1, R2) & fail (R1, R2)                & \cellcolor[HTML]{Feb3ab}fail (R1, R2) & \cellcolor[HTML]{Feb3ab}fail (R1, R2)        & fail (R1, R2)                & \href{https://rawcdn.githack.com/akdimitri/RRAM_COMPILER/24f024d7e8ab4d575474f9ea0d02535fe02467d2/raw/CHARACTERISATION/128_16_16/80_TB_TOP_128_16_16.html}{\textit{link}} \\ \hline
128                       & 32                        & 4                         & pass                                  & \cellcolor[HTML]{Abfeb2}pass & pass                                  & pass                                         & \cellcolor[HTML]{Abfeb2}pass & \href{https://rawcdn.githack.com/akdimitri/RRAM_COMPILER/24f024d7e8ab4d575474f9ea0d02535fe02467d2/raw/CHARACTERISATION/128_32_4/80_TB_TOP_128_32_4.html}{\textit{link}}   \\ \hline
128                       & 32                        & 8                         & pass                                  & fail (R1, R2)                & pass                                  & pass                                         & fail (R1, R2)                & \href{https://rawcdn.githack.com/akdimitri/RRAM_COMPILER/24f024d7e8ab4d575474f9ea0d02535fe02467d2/raw/CHARACTERISATION/128_32_8/80_TB_TOP_128_32_8.html}{\textit{link}}   \\ \hline
128                       & 32                        & 16                        & \cellcolor[HTML]{Feb3ab}fail (R1, R2) & fail (R1, R2)                & \cellcolor[HTML]{Feb3ab}fail (R1, R2) & \cellcolor[HTML]{Feb3ab}fail (R1, R2)        & fail (R1, R2)                & \href{https://rawcdn.githack.com/akdimitri/RRAM_COMPILER/24f024d7e8ab4d575474f9ea0d02535fe02467d2/raw/CHARACTERISATION/128_32_16/80_TB_TOP_128_32_16.html}{\textit{link}} \\ \hline
128                       & 64                        & 4                         & pass                                  & \cellcolor[HTML]{Abfeb2}pass & pass                                  & pass                                         & \cellcolor[HTML]{Abfeb2}pass & \href{https://rawcdn.githack.com/akdimitri/RRAM_COMPILER/24f024d7e8ab4d575474f9ea0d02535fe02467d2/raw/CHARACTERISATION/128_64_4/80_TB_TOP_128_64_4.html}{\textit{link}}   \\ \hline
128                       & 64                        & 8                         & pass                                  & fail (R1, R2)                & pass                                  & pass                                         & fail (R1, R2)                & \href{https://rawcdn.githack.com/akdimitri/RRAM_COMPILER/24f024d7e8ab4d575474f9ea0d02535fe02467d2/raw/CHARACTERISATION/128_64_8/80_TB_TOP_128_64_8.html}{\textit{link}}   \\ \hline
128                       & 64                        & 16                        & \cellcolor[HTML]{Feb3ab}fail (R1, R2) & fail (R1, R2)                & \cellcolor[HTML]{Feb3ab}fail (R1, R2) & \cellcolor[HTML]{Feb3ab}fail (R1, R2) & fail (R1, R2)                & \href{https://rawcdn.githack.com/akdimitri/RRAM_COMPILER/24f024d7e8ab4d575474f9ea0d02535fe02467d2/raw/CHARACTERISATION/128_64_16/80_TB_TOP_128_64_16.html}{\textit{link}} \\ \hline
\end{tabularx}
}
\vspace{-4mm}
\end{table}
}

\begin{table}[!t]
\caption{Memories density comparison.}
\label{tab:comparison}
\vspace{-4mm}
\begin{center}
\begin{tabular}{c|ccc}
\hline
Ref.                           & Feature Size & Technology & Mb/mm\textsuperscript{2} \\ \hline
 %\hline
\cite{miyano2013highly}        & $40\,$nm         & CMOS       & $0.94$     \\
\cite{toh2011characterization} & $45\,$nm         & CMOS       & $0.33$      \\ %\hline
 %\hline
\cite{guthaus2016openram}      & $45\,$nm         & FreePDK45  & $0.826$     \\
\cite{kushida20090}            & $65\,$nm         & CMOS       & $0.77$      \\
%\hline
\cite{lee2019reram}\,$^1$            & $65\,$nm         & CMOS/memristor       & $0.54$      \\
%\hline
This work\,$^2$                 & $180\,$nm       & CMOS/memristor       & $0.024$      \\ %\hline
\cite{antoniadis2021open}\,$^3$   & $180\,$nm       & CMOS/memristor       & $0.082$ \\
\cite{7538307}\,$^3$                 & $180\,$nm       & CMOS       & $0.067$      \\ %\hline
\cite{guthaus2016openram}      & $0.5\,$\textmu m        & SCMOS  & $0.005$     \\ \hline
\end{tabular}
\end{center}
\raggedright\footnotesize{$^1$ RRAM subarray chip. $^2$ Memory chip density. $^3$ Memory cell size.}\\
\vspace{-6mm}
\end{table}

A direct comparison between RRAMs cannot be made as they refer to different sizes and different materials. A density comparison between volatile memories, this work and the RRAM sub-arrays generated by the corresponding RRAM compilers presented in \cite{lee2019reram} and \cite{antoniadis2021open} is shown on Tab.~\ref{tab:comparison}. RRAM compilers generate quite dense designs similar to SRAM designs. This work achieves approximately same density with an SRAM designed at TSMC 180\,nm technology. The memory cell of the proposed implementation is relatively big to ensure robustness and low mismatch. It can be further reduced to improve the density. The optimal size of the RRAM cell should be further investigated. Fully custom made RRAMs exploiting multi state bit memory cells can achieve up to 6.66\,Mb/mm\textsuperscript{2} and read/write access time less than 200\,ns for up to 4\,Mb designs~\cite{sheua4mb2011,sheufast2011,zangenehDesign2014}. The RRAM compiler presented in \cite{lee2021reram} exploits 3D IC design to further improve density, however the results are normalised and a direct comparison cannot be made.

%The density of the generated memories is shown on Tab.~\ref{tab:rram_size}. The RRAM compiler achieves up to 0.024\,Mb/mm\textsuperscript{2} for 180\,nm technology.
%\import{./Tab}{rram_density.tex}
\section{Conclusion} \label{sec:conclusion}

Taking everything into consideration, this paper presented the first open source RRAM compiler which automatically generates the RRAM array, the peripheral circuits and provides layout verification and timing characterisation. In this way, a number of memristors can be integrated and tested by using the automatically generated designs. Finally, the RRAM compiler serves as a great tool for both industry and academia in order to boost the RRAM design procedures and the investigation of the RRAM properties. 
%The RRAM compiler serves as a great tool for both industry and academia as it boosts the RRAM design procedures and offers flexibility on the investigation of RRAM properties. 
%A number of memristors can be integrated and tested by using the automatically generated designs. A number of improvements can be made, though the author suggests that the next step should be the upgrade of the compiler to generate arrays of RRAMs, allowing great scalability.
\section*{Acknowledgment} 
The authors acknowledge the support of the EPSRC FORTE Programme Grant (EP/R024642/1).

\bibliographystyle{IEEEtran}
\bibliography{IEEEabrv,Section/references}

\end{document}